\documentstyle[twocolumn,prl,aps,epsf,graphicx]{revtex}
\begin{document}
\draft
\twocolumn[
\hsize\textwidth\columnwidth\hsize\csname @twocolumnfalse\endcsname
\draft
\title{Saddles in the energy landscape probed by
       supercooled liquids }
\author{
        L.~Angelani$^1$,
        R.~Di Leonardo$^2$,
        G.~Ruocco$^2$,
        A.~Scala$^3$, and 
        F.~Sciortino$^3$
        }
\address{
         $^1$
         Universit\'a di Trento and INFM, I-38050, Povo, Trento, Italy.\\
         $^2$
         Universit\'a di L'Aquila and INFM, I-67100, L'Aquila, Italy. \\
         $^3$
         Universit\'a di Roma {\rm La Sapienza} and INFM, I-00185, Roma, Italy.
        }
\date{MS: LU7493 Revised Version - \today}
\maketitle
\begin{abstract}
We numerically investigate the supercooled dynamics of 
two simple model liquids exploiting the partition of
the multi-dimension configuration space in basins of 
attraction of the stationary points (inherent saddles) 
of the potential energy surface. We find that the 
inherent saddles order and potential energy 
are well defined functions of the temperature $T$. 
Moreover, decreasing $T$, the saddle order vanishes 
at the same temperature ($T_{_{MCT}}$) where the inverse 
diffusivity appears to diverge as a power law. This allows a 
topological interpretation of $T_{_{MCT}}$: it marks the 
transition from a dynamics between basins of saddles 
($T>T_{_{MCT}}$) to a dynamics between 
basins of minima ($T<T_{_{MCT}}$).
\end{abstract}
]

The study of the properties of the  free energy landscape 
or/and of the potential energy  
surface (PES) 
in disordered systems is a topic of current research\cite{tutti}.
The  universality in the dynamic
of systems as different as disordered spin-glasses 
and structural-glasses
\cite{universalita} supports the possibility that some 
universal features at the landscape level 
control the slow-dynamics in these systems\cite{kurch95}.
Along this line, several recent works have attempted to connect both
dynamics and thermodynamics of glass-forming liquids
to landscape properties. 

Numerical investigations of the sampled configuration space
have been performed for several 
models of liquids\cite{nature,heuer,ANGE,barbara,prlentro,thomas}.  
An important outcome of these studies is the demonstration that on cooling
the system populates basins of the PES associated
to local minima 
(the so-called inherent structure $IS$\cite{stillinger}) 
of deeper and deeper depth\cite{nature}. 
The number of distinct basins with the same depth in bulk systems
has also been evaluated\cite{prlentro}. 
These information have
been incorporated in a detailed description of the thermodynamics of
supercooled liquids\cite{heuer,ANGE,prlentro,sri}.

In the landscape framework, 
the dynamics of the system in configuration space is 
conceptually decomposed in a
"fast" oscillatory motion (dynamics within a basin) 
and a slow diffusive motion (dynamics among
different PES basins). 
Quantitative calculation of the
diffusion coefficient $D$ based on landscape properties have
been formulated within the 
Instantaneous Normal Mode (INM) theory\cite{INM}. 
The INM approach focus on the 
properties of the local curvature of the PES sampled by the liquid, 
calculated diagonalizing the Hessian (${\cal H}$) matrix 
of the potential energy $V$. 
Analysis of the resulting eigenvalues and eigenvectors\cite{hessian} 
allows to evaluate 
the number of independent directions in configuration space associated to
basin changes, i.e. to diffusion.
For the cases where such analysis has been
performed, strong evidence has been presented for existence of
proportionality between $D$ and the number of diffusive
directions\cite{st,claudio,wxli,lanave}. It has
also been shown that the number of diffusive directions 
decreases with  $T$ and appears to vanish 
at the so-called Mode-Coupling\cite{mct} transition temperature
$T_{_{MCT}}$\cite{claudio,lanave}, i.e. at the $T$ where an
apparent divergence of the inverse diffusivity is observed.  

A major difficulty in quantitatively pursuing the idea of a connection between diffusivity and topology of configuration space is that no transparent
mapping has been yet proposed to associate equilibrium configurations
to the ``closest'' configuration on the border between different
basins. 

In this Letter we propose such a mapping and 
present an analysis of the properties  of the
closest {\it stationary points} (saddles)
of the potential energy. The proposed mapping, which can be
considered as an extension of the Stillinger-Weber 
mapping\cite{stillinger,quenches},
partitions the configuration space ${\bf R}^{3N}$
of a 3-dimensional $N$ particle system
in basins of attraction of the saddles, characterizing the saddle
with their order and their potential energy. 
The 
dynamics of the system is then described as dynamics between different
saddles' basins. We apply the proposed mapping to two different 
models of simple liquids and  find that: 
(i) the order of the sampled saddle is a well defined function of $T$;
(ii) on cooling  the liquid populates basins associated to saddles of
lower and lower order; 
(iii) the location in potential energy of the saddles is  
much smaller than the system potential energy, providing
 evidence that the diffusion process is entropy driven, 
even below $T_{_{MCT}}$;
(iv) at $T_{_{MCT}}$, the saddle order appears to vanish, 
indicating that, at this $T$,  the system  populates the  basins of 
potential energy minima, confirming that $T_{_{MCT}}$ 
marks the crossover between two different dynamical processes.

To  partition  ${\bf R}^{3N}$ in basins of saddles 
we search for the
basins of attraction of an auxiliary potential function, namely
$W=\frac{1}{2}\vert \vec \nabla V\vert ^2$ \cite{nota2}
(for a similar approach see also \cite{cavagnone}).
The function $W$ is never negative and it is 
zero at all saddle points, i.e. at all points where 
all forces are zero (stationary point configurations).
The saddle points are classified according to their order $n_s$
(the number of negative eigenvalues of $\cal H$ ) 
and their potential energy $e_s$.
Saddles of order zero coincides with 
local minima of the PES (i.e. with IS).
The complete description of the energy landscape would 
require the calculation of the densities of states for each $n$.
However, by investigating
two model systems  we find that
similarly to the case of minima\cite{prlentro},
the saddle's energy and order are well defined function of $T$. 
Hence, 
all relevant information are contained in the functions $e_s(T)$
and $n_s(T)$. 
 
We investigate numerically two simple model liquids: ({\it a})
the monatomic Modified-Lennard-Jones (MLJ) \cite{MLJ}, and, ({\it b})
the standard Lennard-Jones 80/20 binary mixture (BMLJ) \cite{BMLJ}.
Both models are able to support strong supercooling without the
occurrence of crystallization. Standard LJ units are used hereafter.
Equilibrium configurations are prepared by standard microcanonical
molecular dynamics simulations at constant density ($\rho=1$ for MLJ
and $\rho=1.2$ for BMLJ) and at $T$s ranging from the normal
liquid phase ($T \approx 1.6$), down to  $T_{_{MCT}}$.
($T_{_{MCT}}=0.475$ for MLJ and 0.435 for BMLJ).
The systems are composed of $N$=256 (MLJ) and $N$=1000 (BMLJ) particles
enclosed in a cubic box with periodic boundary condition. Truncated
($R_c=2.6$ and $2.5$ respectively)
and shifted LJ potentials are used. We analyze 20 independent 
equilibrium configuration for each $T$.
For each configuration
we calculate the associated IS and inherent saddle
implementing a  steepest descent algorithm which moves
in the direction of $-\vec \nabla V \equiv \vec F$
and $-\vec \nabla W = {\cal H} \cdot \vec F$ respectively 
(the arrows indicates 3$N$-d vectors). 
Finally, the $\cal H$ of the starting equilibrium
configuration (to evaluate  the INM) and of the inherent saddle
(to evaluate $n_s$) is calculated and diagonalised.

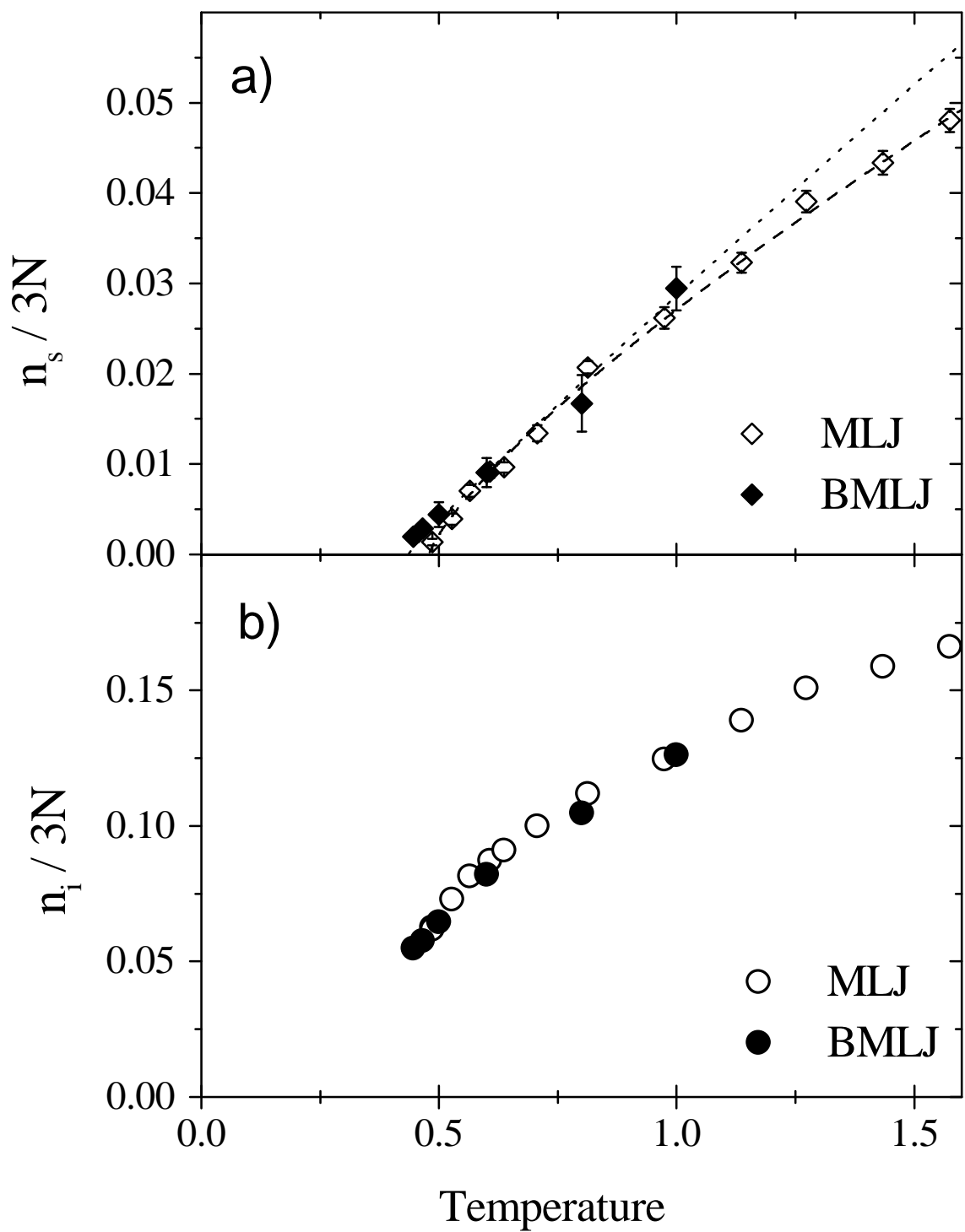
\begin{figure}[t]
\centering
\vspace{-.47cm}
\caption{Temperature dependence of the fraction of the negative eigenvalues of
the Hessian calculated at the inherent saddle configurations $n_s/3N$ (a),
and at the instantaneous configurations, $n_i/3N$ (b).
Open symbols refer to MLJ and full ones to BMLJ.
The dashed line in a) are the best fit to the data with the
function $A(T-T_x)^\gamma$  
(MLJ: $T_x$=0.48$\pm$0.01, $\gamma$=0.78$\pm$0.02.
BMLJ: $T_x$=0.435, $\gamma$=0.94$\pm$0.1.).}
\label{fig1.ps}
\end{figure}

Fig.~\ref{fig1.ps}a  shows the average order of the inherent saddle
(i.e. the number $n_s$ of negative eigenvalues of ${\cal H}$)
as a function of
the equilibrium $T$.  The average
order $n_s(T)$ is a well defined function of
$T$ \cite{nota4}, indicating that the 
trajectory of the system in the configuration space samples statistically
the subspace set up by basins of saddles of a given order $n_s(T)$.
Fig.~\ref{fig1.ps}a shows also that, decreasing $T$, $n_s(T)$
vanishes at $T_{_{MCT}}$. 


This finding gives support to the following scenario: $T_{_{MCT}}$ 
is the $T$ above which the
system explores basins of saddles of order $n_s>0$ and below which 
the system is mostly confined in a local minimum  ($n_s=0$).
The existence of a quantity, $n_s(T)$, that vanishes at $T_{_{MCT}}$
is remarkable. Indeed, this makes $n_s(T)$ a good candidate
for the description of supercooled dynamics, for instance the
computation of $n_s(T)$ is an alternative way of determining $T_{_{MCT}}$.

For comparison with the previous INM studies, 
Fig.~\ref{fig1.ps}b also shows the number of directions 
characterized by negative eigenvalues  $n_i$ as a function
of $T$. As discussed in Ref.~\cite{hessian},  
a non zero value of $n_i$ is found at $T_{_{MCT}}$, 
when the system is trapped in basins of minima, a clear
signature of the presence of non-diffusive unstable modes.
Hence, the introduction of the inherent saddle concept 
offers a way to overcome
the difficulties associated to the presence of
non-diffusive modes in the standard INM approach.
The order of the saddle $n_s$ appears to be a well defined indicator
of the number of diffusive directions.

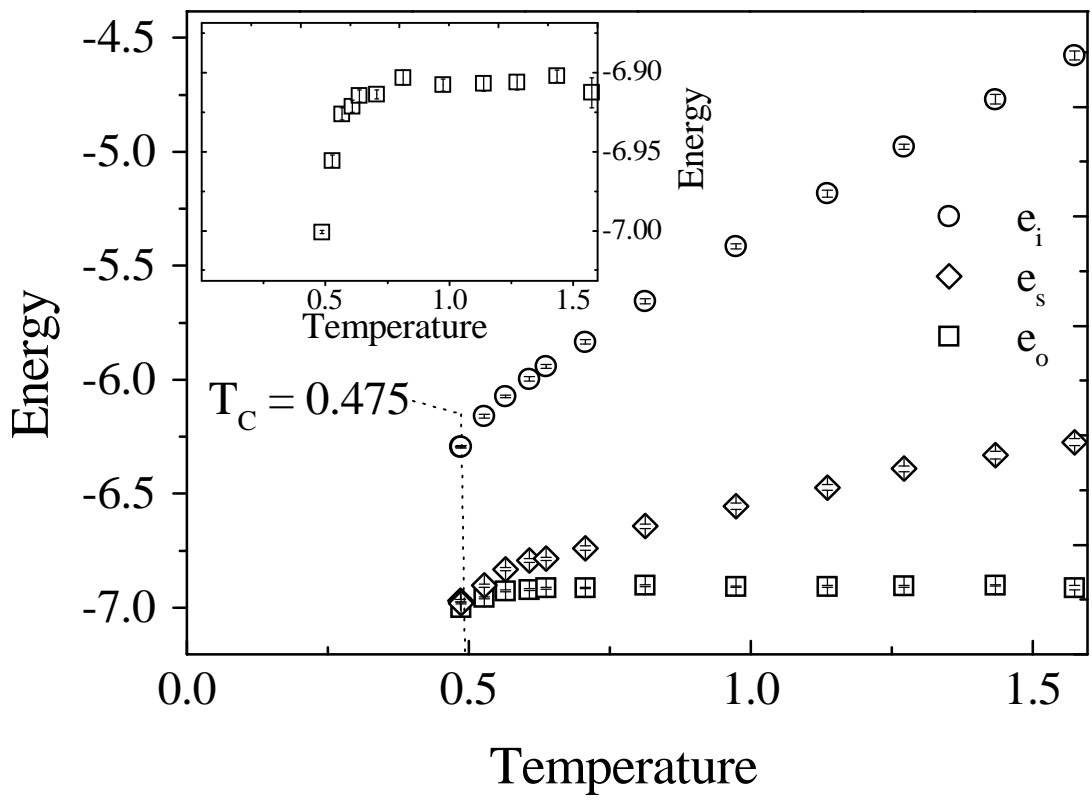
\begin{figure}[t]
\centering
\vspace{-.47cm}
\caption{a) Temperature dependence of the instantaneous energy ($e_i$, $\circ$),
the inherent saddle energy ($e_s$, $\diamond$), and the inherent minima energy
($e_o$, $\Box$) for the MLJ potential model.
The vertical dashed line indicates 
$T_{_{MCT}}$ as derived from the apparent divergence of the inverse diffusivity.
The inset shows, in an enlarged energy scale, the $T$ dependence of $e_o$.
A very similar behavior is found for the BMLJ system (not reported here
for clarity reasons).
}
\label{fig2.ps}
\end{figure}

Next we discuss the location in energy of the inherent saddles
as a function of the equilibrium $T$.
Fig.~\ref{fig2.ps} reports the average instantaneous
potential energy $e_i$, the average potential energy of the
saddle $e_s$ and the average potential energy of the IS $e_o$,
as a function of $T$ for the MLJ potential model. Similar results
holds for the BMLJ case. 
The quantity $e_o$(T) 
 shows a rapid increase between $T_{_{MCT}}$ and $T\approx0.8$
reaching a constant value for higher $T$ (see inset
of Fig.~\ref{fig2.ps} where $e_o$ is reported in an expanded scale). 
We notice that the
overall variation of  $e_o$  is very small on the scale of
the variation of $e_s$ and $e_i$.
The quantity $e_s(T)$ shows an 
intermediate behavior between $e_i$ and $e_o$. 
In agreement with the observation that around $T_{_{MCT}}$ the system explores
basin of attraction of saddles of order zero, we find that 
at $T\sim T_{_{MCT}}$ the inherent saddle energy curve merges on the
IS curve. 
The data reported in Fig.s~\ref{fig1.ps} and \ref{fig2.ps} allow 
to conclude  in an unambiguous way that
$T_{_{MCT}}$ marks the cross-over between two
dynamic regimes: at $T<T_{_{MCT}}$ the system spend most of the time
trapped in a local minima, while at $T>T_{_{MCT}}$ the system explores basins
pertaining to saddle points of increasingly higher -but well defined-
energy and order. This conclusion is consistent with the
INM finding that at $T_{_{MCT}}$ the fraction of  diffusive
directions explored by the liquid goes 
to zero\cite{st,claudio,lanave}. 
It is also consistent with the interpretation of the 
transition between different dynamical regimes at $T_{_{MCT}}$
proposed in Ref.\cite{thomas} and based on the
analysis of the decay of the density-density correlation
functions evaluated along an inherent structure trajectory.
 
An important observation stems from the 
data reported in Fig.s~\ref{fig1.ps} and \ref{fig2.ps}:
the saddles energies are located
well below the instantaneous values.  This finding clearly shows that
the system trajectory is never close to a saddle point
and dynamics should not be described as 
saddle-to-saddle motion, but, more correctly  as dynamics between
basins of attractions of the corresponding saddles.
In this respect, one should not discuss the dynamics of the
system for $T>T_{_{MCT}}$ as activated dynamics \cite{cavagna}.
It is worth to note that the value $e_{i}$  at $T_{_{MCT}}$ 
is larger than $e_s(T)$ for a wide $T$ range.
In other words, even at $T_{_{MCT}}$, the instantaneous potential 
energy is much larger than the energy value at which saddles are located. 
Nevertheless  at $T_{_{MCT}}$  
the system spends a very large fraction of time
in a local minimum and only rarely performs jumps between minima. Hence, 
the diffusion events at low $T$ are
not limited by the presence of energy barrier that must be overcome
by thermal activated processes; they are rather  controlled by the
limited number of directions leading from a basin of a minimum 
to another basin at (almost) constant potential energy.

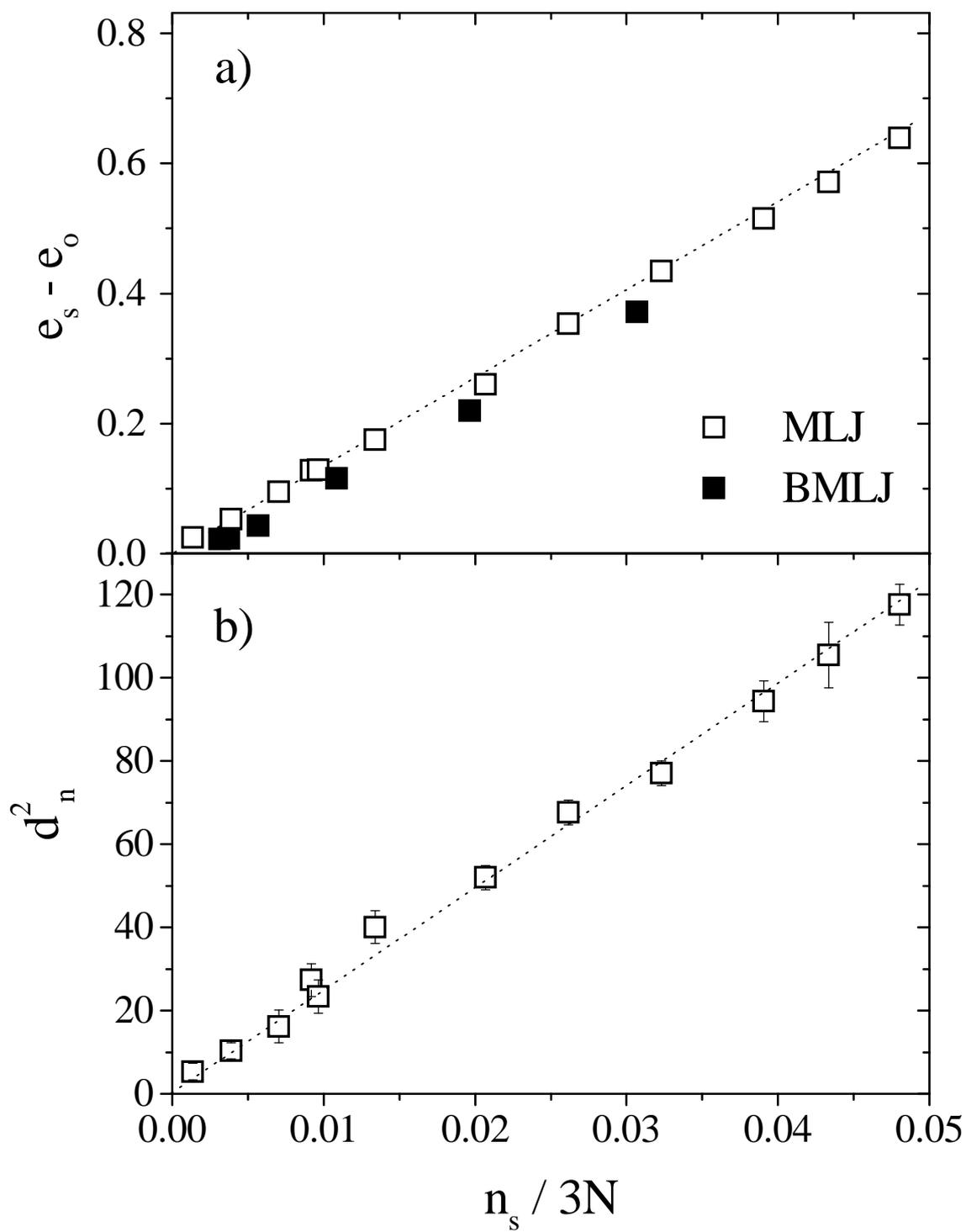
\begin{figure}[t]
\centering
\vspace{-.47cm}
\caption{a) Saddle's elevation energy $e_s-e_o$ as function of the
saddle's order fraction $n_s/3N$ for MLJ (open symbols) and BMLJ (full symbols).
The dashed line, with a slope of $\approx 11$ 
is the best fit to the MLJ data. b) The mean square distance $d_n^2$ between
minima (IS) reached by quenching the inherent saddle of order $n_s$ is plotted
as function of $n_s/3N$ for MLJ. The dashed line is the best fit
to the data}
\label{fig3.ps}
\end{figure}

Fig.~\ref{fig3.ps}a shows the relation between $n_s$ 
and the elevation  (in potential energy) respect to the
corresponding local minima ($e_s-e_o$). We discover a remarkable
linear relationship between these two quantities.
This indicates that -given a minimum-
the energy landscape above it is organized in families of equally
spaced energy saddle points (to rise one step in the saddle order,
the requested energy is $3N\Delta(e_s-e_o)/\Delta n_s \approx 11$
for both systems. Moreover, this simply linear dependence
suggests that the aspects of the landscape above a local minimum
is independent from the energy of the minimum itself.
We also find a linear relation between
the mean square distance $d_n^2$ between minima that have been
reached by quenching the inherent saddles's of order $n_s$ 
and the order of the corresponding saddle\cite{nota5}, as shown in 
Fig.~\ref{fig3.ps}b
This linear relation suggests that 
the descent path from a saddle
of order $n_s$ towards the surrounding minima 
can be interpreted as a sequence of 
independent random steps, each of them decreasing the
order of the saddle by one and increasing the (squared) distance
between the associated local minima by a fixed amount
(random walk among saddle points).
It is worth to note that these properties of the energy
landscape (i.e. the linear dependence between $e_s-e_o$,
$d^2_n$, and $n_s$) are ingredients of a model for the landscape
introduced by Madan and Keyes \cite{keyes93} and recently revisited
\cite{keyes00}. This similarity deserves
further investigation.

In conclusion, by the numerical investigation of two different systems,
we have pointed out the relevance of the concept of inherent saddle
in describing the dynamics of supercooled liquids  and
we have highlighted same important characteristics of the energy
landscape. In particular we have: {\it i)} shown that the inherent
saddles' properties, energies and orders, are well defined function
of $T$; {\it ii)} demonstrated the validity of the conjecture that
$T_{_{MCT}}$ marks the transition between a dynamic among minima
($T<T_{_{MCT}}$) and a dynamics where the systems spend time nearby high
order stationary points; {\it iii)} found that $n_s(T)$ is a quantity
that can be efficiently used to measure numerically $T_{_{MCT}}$; 
{\it iv)} provided evidence that the diffusion
processes are entropy driven; and {\it v)} shown that the aspect of
the energy landscape ``seen'' by a given local minimum is highly
regular (as demonstrated by the linear dependence of $e_s-e_o$ and
of $d_n^2$ from $n_s$) and independent from the specific minimum.

We acknowledge partial support from MURST-PRIN-98.
FS acknowledges partial supports also from INFM-PRA-HOP.


\begin{references}
\bibitem{hessian}
In the INM approach, 
the positive eigenvalues of the Hessian (${\cal H}$) 
are associated to vibration around quasi-equilibrium position,
while the modes with negative eigenvalues are assumed to contain all
the information relevant to the description of the long-time
(diffusive) dynamical processes.  It is not obvious that a negative
curvature in ${\cal H}$ is necessarily a diffusion coordinate in
configuration space; for example, in the crystal phase --- thought the
diffusion is absent --- negative curvatures are present at finite
$T$\protect\cite{anarmon}.  One finds that non-diffusive unstable
modes are present even in the liquid and glassy phase at all
$T$\protect\cite{BEMBE}.  This makes the identification of those
unstable modes which really contribute to diffusion the main
difficulty of a straightforward INM approach\protect\cite{berne,claudio}.
Indeed, model-dependent recipes are requested to sort-out the
diffusive directions (i.e. the direction which leads to distinct
basins) from the body of directions characterized by negative
eigenvalues.
\bibitem{nota2}
It is worth to note that for every stationary point of $V$, $W$ has
a minimum (actually $W=0$ there), while the reverse is not true.
There are minima of $W$ that are not stationary point of $V$
(flex points). These points can be easily identified as 
the Hessian has one or more eigenvalue(s) much smaller than
those found in the ``true'' saddle points.
 We find that the number of the flex points is negligible
 with respect to the number of true saddles of $V$
and their existence introduces an error
of approximately 2\% in the estimation  of $e_s$ and $n_s$.
\bibitem{nota4} 
This is evident by the observation that the distribution
functions $P(n,T)$ are very narrows and their variance become
smaller and smaller increasing $N$.
\bibitem{nota5}
Starting from a given saddle point, we randomly perturb the particle 
coordinates, then we start the minimization of $V$
from these perturbed structures. The 3N-D euclidean
distance between the so found minima configurations is then calculated.
\end{references}
\end{document}